\documentclass[12pt]{article}
\usepackage{dcolumn}
\usepackage[ansinew]{inputenc}
\usepackage{siunitx, tabularx, etoolbox, booktabs}
\newcommand{\ubold}{\fontseries{b}\selectfont}
\robustify\ubold
\usepackage{booktabs}
\usepackage[T1]{fontenc}
\usepackage[skip=2pt,font=scriptsize]{caption}
\RequirePackage{amsthm,amsmath,amssymb,amsthm}
\RequirePackage[round]{natbib}
\RequirePackage[colorlinks,citecolor=blue,urlcolor=blue]{hyperref}
\RequirePackage{hypernat}
\RequirePackage{enumerate}
\RequirePackage{enumitem}
\RequirePackage{graphicx}
\RequirePackage[usenames,dvipsnames]{xcolor}
\RequirePackage[mathscr]{eucal}
\usepackage{float}
\usepackage{tikz}
\usepackage{setspace}
\usepackage{verbatim}
\usepackage{mathtools}
\usepackage[round]{natbib}
\usepackage{enumerate}
 \usepackage{url}
 \usepackage{adjustbox}

\usepackage{marvosym}
\usepackage{url}

\usepackage{caption,newfloat}

\DeclareMathOperator{\E}{\mathbb{E}}
\usepackage[margin=1in]{geometry}
\usetikzlibrary{matrix}
\usepackage{expl3}
\ExplSyntaxOn
\cs_new_eq:NN \Repeat \prg_replicate:nn
\ExplSyntaxOff
\usepackage{verbatim}
\tikzset{
    vertex/.style = {
        circle,
        fill            = black,
        outer sep = 2pt,
        inner sep = 1pt,
    }
}

\DeclareCaptionType{ Box}  
\usepackage[]{algorithm2e}
\begin{document}

\begin{titlepage}
\title{Testing the Presence of Implicit Hiring Quotas with Application
to German Universities}
\author{ Lena Janys\thanks{University of Bonn (Department of Economics), Hausdorff Center of Mathematics and IZA} \thanks{I thank Christina Felfe, Joachim Freyberger, Guido Friebel, Hans-Martin von Gaudecker, Daniela Grunow, Astrid Kunze, Shelly Lundberg, Pia Pinger, Jesper Roine, Amelie Schiprowski, Christopher Walsh and Lingwei Wu, as well as participants of the ``CEPR Women in Economics workshop'', the workshop ``The Determinants of Careers for Women and Men: Institutions, Professions, Organizations and Households'' and the ``Forum for Research on Gender Economics (FROGEE) webinar'' for helpful comments and suggestions. The the R files to calculate all quantities is available on my GitHub-page \url{https://github.com/LJanys/Implicit-quota-package}.}}

\maketitle


\begin{abstract}


Women are underrepresented in academia in general and economics in particular. I introduce a test to detect an under-researched form of hiring bias: implicit quotas. I derive a test under the null hypothesis of gender-blind hiring that requires no additional information about individual hires and can be used to analyze hiring bias in a variety of other hiring settings. I derive its asymptotic distribution and propose a parametric bootstrap procedure that resamples from the exact distribution. I analyze the distribution of female professors at German universities and find an implicit quota of one or two women on the department level.  
  

\end{abstract}

\end{titlepage}

\doublespacing
\section{Introduction}\label{sec:intro}


Although the share of women among university graduates and women's labor market participation has increased dramatically, the share of women among high-ranking professionals has remained low. For example, in 2019, women accounted for just 27.8\% of board members of the largest publicly-listed companies registered in EU countries, and only about 17\% of senior executives are female (see (\cite{eurostat2019}). In Germany, the share of female board members among the 30 largest publicly-traded companies in the DAX even fell and is currently at only 12.5\% (\cite{allbright2020}).\footnote{This extends to other higher-level professions. For example, women hold only 32\% of seats in European parliaments, and there were only four female heads of government (15\%) in 2019 (see \cite{eurostat20192}).}  

Understanding the role that candidate gender plays in hiring is crucial if one wants to increase the representation of women. One possible channel through which gender may influence hiring decisions might be implicit quotas. Implicit quotas, similarly to explicit quotas, are quantity restrictions on hiring but without explicit targets. They can be seen as a form of discrimination, as candidate gender is taken into account in the hiring decision, both positively and negatively. However, proving discrimination in hiring is difficult, as researchers rarely observe enough information on hiring committees and candidates to make definitive judgments (see for example \cite{Arciharvardnber}). To detect discrimination against women in academia, we would need to have detailed information on individual candidates' qualifications, the preferences and information set of the hiring committee, and the number of women already in the department. 

In this paper, I introduce a statistical test to detect implicit quotas. This test requires relatively little information in order to make statements about possible hiring bias. I develop a test for gender-blind hiring with implicit quotas as alternatives. I characterize both the small sample distribution, which can be approximated numerically, and the asymptotic distribution under the null of gender-blind hiring. In my application I study the distribution of women in German universities. 

My main results show that (1) the current allocation of women across departments is highly improbable under the null hypothesis where candidates are drawn gender-blind from a Bernoulli distribution with the share of female professors $p$. Specifically, there are ``too few'' departments with no female professors and ``too many'' with one or two female professors across disciplines. This result holds both using the exact distribution which can only approximated numerically, and the critical values using the asymptotic distribution. (2) I find no evidence that disciplines with a higher share of women are different in their implicit quotas, and I can show that the distribution of female shares across \emph{disciplines} could be well-explained by a two-women quota per department in all disciplines. This test can be used in many different settings where one can reasonably assume that hiring probabilities for underrepresented groups are approximately the same across units, such as management levels across different firm locations or on the partner-level in top law firms. 

German universities are ideal for studying this topic. Germany has a homogenous nationwide university system that is entirely publicly funded, and almost all universities offer a wide array of disciplines. This homogeneity is essential to identifying potential implicit quotas because the assumptions, especially an approximately constant hiring probability within disciplines, are more likely to be met.

This paper adds to several strands of literature on the representation of women in top positions. One policy measure which has -- by design -- significant effects on the representation of women in the short run is \emph{explicit} quotas. However, even in the absence of explicit quotas, there may still be quantity restrictions on diversity, what I call \emph{implicit quotas} in the paper. Donors, politicians, NGOs, and society put pressure on organizations and institutions to commit to increasing the representation of women. Crucially, however, these pressures rarely have explicit targets. Evidence on implicit quotas is sparse, with the notable exception of \cite{chang2019diversity}, who study female representation in boards and find evidence for an implicit quota on the boards of the largest publicly listed companies in America.



There is a large and growing literature showing positive effects in that explicit quotas raise female participation. Explicit quotas have, for example, been analyzed in the context of company boards or election lists, see for example \cite{bertrand2018breaking} evaluating a Norwegian board reform, and \cite{maida2019female} on an Italian board reform. \cite{bagues2021can} analyze a Spanish quota where at least 40\% of candidates on ballots were mandated to be of either gender. \cite{balafoutas2012affirmative} show that explicit quotas increase willingness for competition without decreasing group performance. However, evidence of other potential benefits of explicit quotas, such as firm performance or potential spillover effects on lower-level managerial or electoral positions, is mixed. Additionally, there are some significant drawbacks associated with explicit quotas. Talent might be misallocated, which would harm both employers, as well as female professionals who will not necessarily be matched with an employer where they can develop their full potential, as pointed out by \cite{bagues2021can}.

 Diversity pressure, as an alternative to explicit quotas, seems to be quite common: many political institutions, professional organizations, and firms explicitly mention that they want to increase the share of women. Universities are particularly prone to implicit quotas, as the share of female professors is low (see, e.g., \cite{lundberg2020} and \cite{friebel2019women}), and increasing the share of women via explicit quotas is, for legal and political reasons, difficult to implement.\footnote{See for example the recent ruling against the Eindhoven University of Technology's gender quota program by the dutch human rights commission, \cite{eindhoven2020}.} A low share of female professors in universities probably has substantial costs for society. First,  male professors might focus on different content and methods. Second, many young professionals study and receive their education from professors in universities, which might affect many decisions across their professional life. Related, female professors provide role models, which may affect the decisions of young female (but also male) professionals. For example, female role models might affect educational choices and aspirations (\cite{beaman2012female}), specialization choices (\cite{porter2020gender}, \cite{buser2014gender}), labor supply in later life, or even the allocation of paid and unpaid labor in households. Unsurprisingly, politicians, student representations, donors, third-party funding agencies, and the general public pressure universities to increase the share of women.


My results show that politicians and donors have to be aware of potential implicit quotas that could become entrenched without the accountability of an explicit target. This likely requires different monitoring to detect whether the shares are increasing beyond certain thresholds. While this analysis is not dynamic and does not predict whether in a few years implicit quotas will increase too, e.g., to 3-4 women, we can make inferences based on another statistic: the implicit quota is centered around the \emph{number} of women in the department, \emph{not} around the discipline mean. This implies that administrative pressures to conform to a particular number of women are on large and small departments alike. 


In the presence of implicit quotas, it is doubtful whether analyzing productivity indicators, such as publications and citations is necessarily informative for two reasons. (1) match quality might be worse for women, as the probability of being hired at a particular institution relies on factors outside their control, namely on the sum of the women already at the department. (2) if women are more likely to be hired when there are no women but less likely to be hired when there are already two, then the characteristics of the third woman in a department will potentially be very different from those of the first one. Hence, analyses that analyze productivity patterns of scientists in different disciplines, such as \cite{huang2020historical}, will mask significant heterogeneity.\footnote{In fact, one of their key findings, that research productivity differences have increased over time could at least partially be explained by an increase in quality heterogeneity.}


 The presence of implicit quotas can also be informative about how employers, the public, or other stakeholders perceive diversity. There is also evidence that implicit quotas can be self-reinforcing. In a recent experimental paper, \cite{paryavi2019descriptive} analyze how descriptive norms in gender composition are acted upon by men and find that descriptive norms do not lead to prescriptive norms and can even lead to backlash if male ``employers'' are informed that others have hired more women. Suppose tokenism plays a role and implicit quotas result from a shift in social pressure (either by the university administration or funding agencies). In that case, it is unclear whether the ``acceptable number'' of women will increase over time or whether there will be stagnation since there is no explicit target.

This paper proceeds as follows: In Section (\ref{sec:data}) I describe the data and the institutional background in Germany. Section (\ref{sec:method}) introduces the method. In Section (\ref{sec:test}) I develop the test statistic, in Section (\ref{sec:as_test}) I derive the asymptotic normality result for the test statistic and in Section (\ref{sec:boot}) I introduce the parametric bootstrap procedure. Section (\ref{sec:app}) presents the results from the application. Section (\ref{sec:disc}) concludes and discusses some implications.

\doublespacing
\section{Data and institutional background}\label{sec:data}

In this section I will briefly describe the general institutional setup in Germany and the data I use. 
\subsection{Universities in Germany}

There are about 70 Universities in Germany, of which most offer the full spectrum of study disciplines. There are 17 Universities of Technologies more tailored towards the natural sciences and engineering but employ faculty in the social sciences and humanities. This feature of the German academic system allows me to identify implicit quotas since there needs to be ``enough'' observed variation across departments and (relatively) homogenous hiring criteria across universities. In general, the ``Fachbereich'' (department) makes official hiring decisions. Disciplines are organized in so-called faculties (``Fakult\"aten''). The associated faculty-council forms a hiring committee, typically comprised of professors within a discipline. This hiring committee screens applicants and makes a formal hiring recommendation in the form of a (ranked) list with one or more suitable candidates to the official faculty council. If the faculty council approves the list, it is put up to a vote to the academic senate of the entire university and then signed by the university's president. In practice, the list recommendations of the hiring committee are rarely overturned at subsequent stages. However, the diversity incentives and the average make-up of the hiring committees make me confident that the actual hiring decision is taken on the discipline level. In Germany there are no explicit quotas for professors, but there is a lot of diversity pressure from different institutions. Universities have to report the share of women in professorial and academic positions regularly to the government. The German Research Foundation (DFG) requests that in all funding applications, regardless of discipline, applicants report the gender dimensions of the project. In the case of large funding requests, applicants have to state the share of women who are participating and explicitly outline strategies to increase diversity. The pressure is also reflected in job ads by German universities: Almost all job ads for professorships in Germany explicitly mention that universities have the goal to increase the representation of women and mention that they explicitly invite applications from female applicants.

\subsection{Data and Definitions}\label{sec:def}

I examine the distribution of female faculty in all German universities using administrative data from the German Federal Statistical Office (DeStatis) in 2015. This data contains information on all non-administrative employees at German universities.\footnote{``Personalstatistik: Wissenschaftliches Personal''} For professors, it contains, among other things, information on the discipline, their gender, their salary category, their university, the year of their first appointment as a professor, and their year of birth.

\paragraph{University}

I only consider universities and not universities of applied sciences because these do not overlap in their hiring pool and have different populations of students and working conditions.\footnote{Note that the official statistics that are published on the official materials by DeStatis \emph{do not} contain information on university status, so I coded status as a university myself. This re-coding was unproblematic since the actual names of the universities are in the data and status is unambiguous, but it does mean that some officially published statistics are not directly comparable.} 

\paragraph{Professors}
According to the administrative classification, I restrict the sample to professors, both tenured and untenured, including assistant- associate and full professor level, but not substitute professors. I do not condition on civil servant status or full-time employment, although neither of these would change the sample in a significant way. 

\paragraph{Discipline}
I define discipline according to the administrative three-digit classification of the statistical office.\footnote{ ``Schlüsselverzeichnisse für die Personalstatistik'', which is also used to define disciplines used for allocating third-party funding at the German Research Foundation (DFG).} The list of disciplines is given in Table (\ref{table_sd_llo}) below. I only consider disciplines with at least three departments.
Some disciplines did not meet this minimum threshold and were excluded.

\paragraph{Department}
 
 I define a department as all faculty that are employed within a discipline at the same university, as there is no direct department designator in the administrative data. 
However, professorships are allocated according to a so-called position plan (``Stellenplan''). This means that the ministry of culture in each state allocates the number of professorships to universities and the respective teaching and research units (so-called ``Lehr- und Forschungsbereiche''). The discipline code that I use here corresponds to these ``Lehr- and Forschungsbereiche''. Universities cannot independently create professorships without approval. Once a position is created, a specific teaching load (depending on the state and the seniority of the position, i.e., Assistant/Associate/Full Professor) is allocated to this ``Lehr- und Forschungsbereich'', e.g., economics. Hence, these organizational units are then responsible for this increase in the overall teaching load, and there is no incentive to ``give away'' this position to a different organizational unit, i.e., department. Furthermore, what I observe here is the allocation of the \emph{position} to the teaching and research unit (e.g., economics), not the actual classification of the researcher. Therefore the unit the position is allocated to is most likely also the unit that makes the hiring decision.


\bigskip

\begin{figure}[h!]
\hspace{-1cm}
	\input{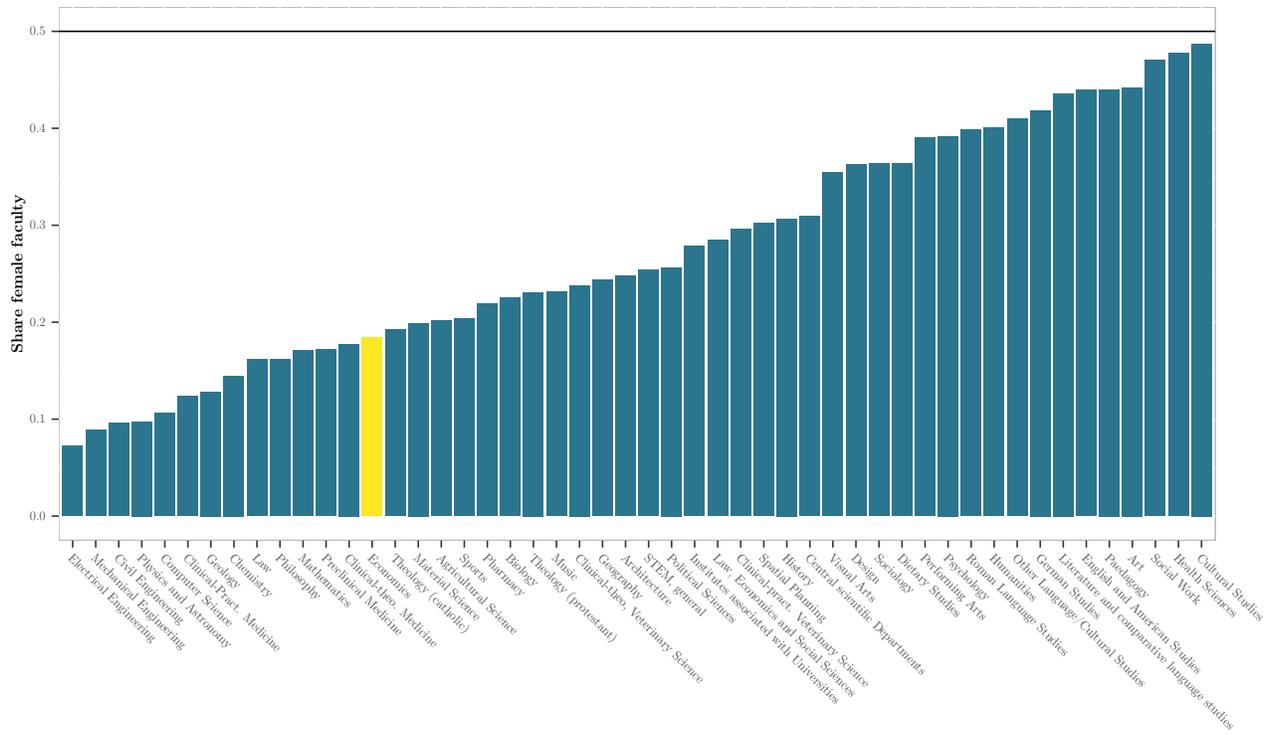}
	\caption{Share of female faculty by discipline, ordered, economics is highlighted for reference. The black horizontal line indicates 0.5, i.e. there is no discipline where women hold a majority of professorships.}\label{fig:fem_share_econ}

\end{figure}

Figure (\ref{fig:fem_share_econ}) shows the (ordered) shares of female professors by discipline. Economics is highlighted for reference. The black horizontal line indicates 0.5, i.e., there is no discipline where women hold most professorships. The final sample contains $n=1737$ departments across 50 disciplines. 

\newpage

\newpage

\doublespacing
\section{Method}\label{sec:method}

Analyzing implicit quotas is difficult: First, whether or not hiring is gender blind is in most cases unobservable, and the decision rule used by employers can only be inferred from observed outcomes using mostly unobservable individual characteristics. (see e.g. \cite{Arciharvardnber}).
Second, because the implicit quota itself is not known, the probability of hiring a woman is not \emph{uniformly} lower given the same observable characteristics. Here, I exploit the fact that many departments at different universities offer the same discipline and thus hire from the same hiring pool. Let $s \in \{1,...,S \}$ be an index for the discipline. For a given probability of hiring a woman $p_s$ in discipline $s$ -- which I treat as known and which corresponds to the observed share of women within a discipline (see Figure (\ref{fig:fem_share_econ}))-- the distribution of women across departments should follow a known distribution. Intuitively, if the number of departments is large enough, even for $p_s$ quite low or quite high, we would expect ``tail events'' such as departments with zero women or departments with many women to occur. The challenge then becomes how to quantify and formally argue what an ``unusual distribution'' looks like.


 \subsection{Testable predictions}\label{sec:test}   
 
  If the distribution of female faculty across departments is gender-blind and depends only on the overall discipline probability $p_s$, I would like to make testable statements about the deviations of the observed number of departments with $z$ female faculty from their expected value. Let $z$ denote the number of women in each department, e.g., $z=0$ denotes zero women department, $z=1$ denotes one-woman department.
  
  We can formalize this by thinking about each employment decision as a separate Bernoulli experiment with associated success probability $X_i \sim Bernoulli(p_s)$. The success probability corresponds to the average share of women in the discipline, so that every employment decision is like tossing a coin with the outcome ``woman is hired'' occurring with probability $p_s$, which I treat as known.
   

 \paragraph{The null hypothesis}
  
Let $n$ denote the number of departments over all disciplines. Let $n_s$ denote the number of professorships in discipline $s$, for $s=1,...,S$. I want to test the joint null hypothesis whether the hiring process is random (i.e. gender-blind) across all departments. In terms of this application, this means that the null hypothesis $$
H_{0}: {X}_{i} \overset{i.i.d.}{\sim} {Bernoulli}(p_s) \, \text { for each } i \text { in } 1, \ldots, n_s \text{ and } s=1,...,S
$$
assesses whether the hiring process is gender-blind for all departments over all disciplines. 

Let $d$ be the index for each department, such that $d=1,.., n$. Let $n_{d}$ be the number of department members of department $d\in \{1,...,n\}$. For each department $d$, the number of women is given by
\begin{equation}\label{eq:y}
	Y_{d}=\sum_{i=1}^{n_{d}}X_i
\end{equation}

 Under the Null of gender-blind hiring, the number of female professors in each department is $Y_{d}\overset{independent}{\sim} Bin(n_{d},p_{d|s})$, where $p_{d|s}$ is the $p_s$ for discipline $s$ that department $d$ belongs to. Thus, we can calculate for each department the probability of observing exactly $z$ women explicitly as 

\begin{equation}\label{eq:pzds}
	p_{d}(z):=P\left(Y_{d}=z \right) = \left(\begin{array}{l}n_{d}\\ z\end{array}\right)p_{d|s}^z(1-p_{d|s})^{n_{d}-z} , \,\,  z=0,1,2,...n_d
\end{equation}

For every department, I define the random variable

\begin{equation}\label{eq:hd}
	H_{d}(z)=\mathbf{1}\left(Y_{d}=z\right)
\end{equation}

under the Null of gender-blind hiring it follows directly that $H_{d}(z)\overset{independent}{\sim} Bernoulli\,(p_d(z))$ for $z=0,1,...,n_d$.

To make testable statements about the sum over all departments, I need to know the distribution of the sum of the $H_d(z)$ over $d$, i.e. $\sum_{d=1}^n H_d(z)$. If $p_d=p$, i.e. if the success probabilities for each department were identical, the sum would follow a binomial distribution. However, because this is not the case, the sum of all $H_d(z)$ follows a so-called poisson binomial distribution with mean $\mu_n=\sum_{d=1}^np_d(z)$ and variance $ s_n^{2}=\sum_{d=1}^n\sigma^2_d$ with $\sigma^{2}_d =\left(1-p_{d}(z)\right) p_{d(z)}$, see also \cite{hong2013computing}. Below I will show that this distribution converges to a normal distribution as $n \rightarrow \infty$. There exist other approximations for Poisson binomial random variables, and if $n$ is very small, it might be feasible to calculate exact expressions of the mean and variance based on convolution, see \citep{liu2018approximating}. However, we are ultimately interested in the distribution of the sum across many different disciplines, and it is not clear that any reasonable closed-form solution exists in that case.

 I am interested in alternatives that imply implicit quotas at specific values of $z$. This entails that the observed realizations differ systematically across departments from their expected value under the Null. I formulate the following test statistic for different specific values $z$, where $z$ indicates the number of women in the department
\begin{equation}\label{eq:Hz}
	{T_n}(z)=\frac{\sum_{d=1}^{n}H_{ d}(z)-p_d(z)}{\sqrt{\sum_{d=1}^{n} \sigma_{d}^{2}(z)}}
\end{equation}

For the joint null hypothesis of gender-blind hiring across departments I want to quantify the uncertainty around the sum of the deviations.

\subsection{Asymptotic distribution of the test statistic}\label{sec:as_test}



Let $B_{ d}(z)=H_{d}(z)-p_{d}(z)$ be a sequence of centered, independent Bernoulli random variables and $s_{n}^{2}=\operatorname{Var} (\sum_{d=1}^{n}B_{d})=\sum_{d=1}^{n} \sigma_{d}^{2}.$ Even though the distributions of the $B_{ d}(z)$ do not depend on $n$ if the $n_d$ are fixed, we could generalize this setup to $B_{n, d}(z)$ as a row-wise triangular array of independent Bernoulli random variables. This would be the case if, for example, the $n_d$ could vary. I will use this more general set-up in the subsequent asymptotic analysis, although here I treat the $n_d$ as fixed. Clearly, $\sum_{d=1}^{n}B_{n, d} / s_{n}^2$ has mean 0 and variance $1$ if $H_0$ is true. $\hat{T}_{n}(z)=\frac{\sum_{d=1}^{n}B_{n ,d}}{s_{n}^{2}}$ given by \eqref{eq:Hz}, is then asymptotically standard normal with limiting distribution given by

\begin{equation}\label{eq:h_dist}
	\hat{T}_{n}(z) \stackrel{d}{\rightarrow} N\left(0, 1\right), \, \, \, \text{for} \,\, n\rightarrow \infty 
\end{equation}

It is sufficient to show that the Lyapunov condition holds to ensure \eqref{eq:h_dist} (see for example Theorem 2.7.1 \cite{lehmann2004elements}).

\paragraph{Lyapunov condition} 


	Let $\{W_{n,d}\}$ be a row-wise independent triangular array of random variables, with $\E (W_{n,d})=0$ and $\E\left(W_{n,d}^{2}\right)=\xi_{n,d}^{2}$, for $d=1,...,n$ and $s_n^2=\sum_{d=1}^n\xi_{n,d}^{2}$.

	\begin{center}
 There exists $\delta>0$ such that $\frac{1}{s_{n}^{2+\delta}} \sum_{d=1}^{n} \E\left(\left|W_{n,d}\right|^{2+\delta}\right) \rightarrow 0$ as $n \rightarrow \infty$.
\end{center}

If the Lyapunov condition holds, then $\hat{T}_n$ converges in distribution as in $\eqref{eq:h_dist}$.
It was established above that $\E (B_{n,d})=0$ and  $\E\left(B_{n,d}^{2}\right)=\sigma_{nd}^{2}.$ It holds that for any $\delta>0$,
$$
1 \geq \underbrace{p_{d}(z)\left(1-p_{d}(z)\right)}_{\sigma_{d}^{2}}=\E\left( B_{ n,d}^{2}\right) \geq \E\left(\left|B_{ n,d}\right|^{2+\delta}\right)
$$
Therefore,
$$
\frac{1}{s_{n}^{2+\delta}} \sum_{k=1}^{n} \E\left|B_{ n,d}\right|^{2+\delta} \leq \frac{1}{s_{n}^{2+\delta}} \sum_{d=1}^{n} \operatorname{Var} (B_{n,d})=\frac{1}{s_{n}^{\delta}}
$$
Therefore, if $s_{n} \rightarrow \infty$, which is true as long as $p_{d}(z)$ is bounded away from 0 and 1, the Lyapunov condition is satisfied and I can conclude that $\sum_{d=1}^{n} B_{ n,d} / s_{n} \stackrel{{d}}{\rightarrow} N(0,1)$.\footnote{As a technical aside, if the distribution of $B_{n,d}$ actually depends on $n$ though varying $n_d$, this condition is only ensured if the $n_d$ are bounded in some way.}

Rejection of the null hypothesis indicates that the \emph{joint} Null of gender-blind hiring across all departments is rejected, but does not indicate \emph{which} departments or disciplines engaged in non-gender blind hiring.

\bigskip

 For the normal approximation to work well in this context we thus only need to assume that $n$ is large, which means that either the number of departments within disciplines and/or the number of disciplines are large, but we cannot make statements about particular departments or disciplines.

\subsection{Parametric Bootstrap}\label{sec:boot}

As an alternative to the asymptotic test procedure, I introduce a type of parametric bootstrap procedure, or more precisely a numerical approximation of the exact distribution. I first draw a random value from a Bernoulli distribution for each position $X_i$, with the $p_s$ from the data and calculate $H_{d}(z)$ for all $d=1,...,n$ (collected in a vector $\mathbf{H(z)}$) according to \eqref{eq:hd} with $n=1737$ from the data. I center the resulting data using $p_d(z)$ for each $d$ and calculate the sum over all departments $B(z)=\sum_{d=1}^nB_d(z)$. I treat the resulting value as one bootstrap observation. I denote the vector of department sizes $\mathbf{n^d}=\left(\begin{matrix}
	n_1&n_2&\cdots&n_n
\end{matrix}\right)$.

In detail, the procedure is the following. Let $B$ be the number of bootstrap draws and $^*$ denote a bootstrap sample. 

%

%

\IncMargin{1em}
\begin{algorithm}
\SetKwData{Left}{left}\SetKwData{This}{this}\SetKwData{Up}{up}
\SetKwFunction{Union}{Union}\SetKwFunction{FindCompress}{FindCompress}
\SetKwInOut{Input}{input}\SetKwInOut{Output}{output}

\BlankLine
\For{$z=0,...,10$, $b=1^*,...,B^*$}{\BlankLine Draw each position $X_{i}$ for $i=1,...,n_s$ over all disciplines $s=1,...,S$ from a $Bernoulli(p_s)$ and collect these in a vector denoted by $\mathbf{X^{b^*}}$. 
\BlankLine
\For{For $d=1,...,n$}{calculate $Y_{d}^{b^*}$ (according to \eqref{eq:y}) using the vector of department sizes $\mathbf{n^d}$ from the data. 
Calculate $B_d(z)^{b^*}$ as described above.
\BlankLine }
Calculate $B(z)^{b*}=\sum_{d=1}^n B_d(z)^{b^*}$ and compare the empirical distribution of the Bootstrap realizations $B(z)^{b^*}$ with the actual ${B}(z)$ from the data. 
}
\end{algorithm}\DecMargin{1em}

\doublespacing
\section{Results}\label{sec:app}

\subsection{Descriptives}

\paragraph{Share of female professors across disciplines} The mean share of female professors varies significantly across disciplines. Traditional STEM subjects, law, and economics have a relatively low share of female professors, ranging from about 7~\% in electrical engineering, 16~\% in mathematics, and (18~\%) in economics. The share is higher in other disciplines in the social sciences or the humanities. For example,  political science has a share of 26~\%; history has 30~\% and german (40~\%). However, it is notable that there is no discipline where women hold a majority of positions: the highest share of female professors is in cultural studies, where women hold about 48~\% of positions. 
In 2015, 50~\% of male employed professors were appointed in 2004 or after, while 50~\% of currently employed female professors were appointed in 2008 or after.

 %



\paragraph{Average department size}

Departments in the humanities, where the average share of women is higher than in STEM subjects, law,  and economics, tend to be much smaller. In cultural studies, the average department only has six professors, where the average electrical engineering department has about 19, and the average economics department has about 22. Again, the other social sciences are in between. This is broadly consistent across countries, e.g., in 2010, the average physics department employed 29.2 full-time faculty vs. history, which had an average department size of 16.5 in 2007 among Ph.D. granting institutions.\footnote{see \cite{physics} and \cite{history}.}

%

\subsection{Testing for implicit quotas}
Here I present the deviations from the expected value as outlined above, both overall and by discipline. To characterize whether the distribution of women is unusual and violates the null of gender-blind hiring, I will present both the result from calculating the value of the test statistic \eqref{eq:Hz} and the critical values from a two sided test using the standard normal distribution and the quantiles from the parametric bootstrap procedure. Overall, both procedures clearly indicate that there are too many departments with exactly two women and too few with zero-, and three, four or five women. Based on this evidence, I can reject the joint Null of random hiring with a similar p-value for both the normal approximation and the parametric bootstrap procedure.

\subsection{Main results}\label{sec:main_res} 
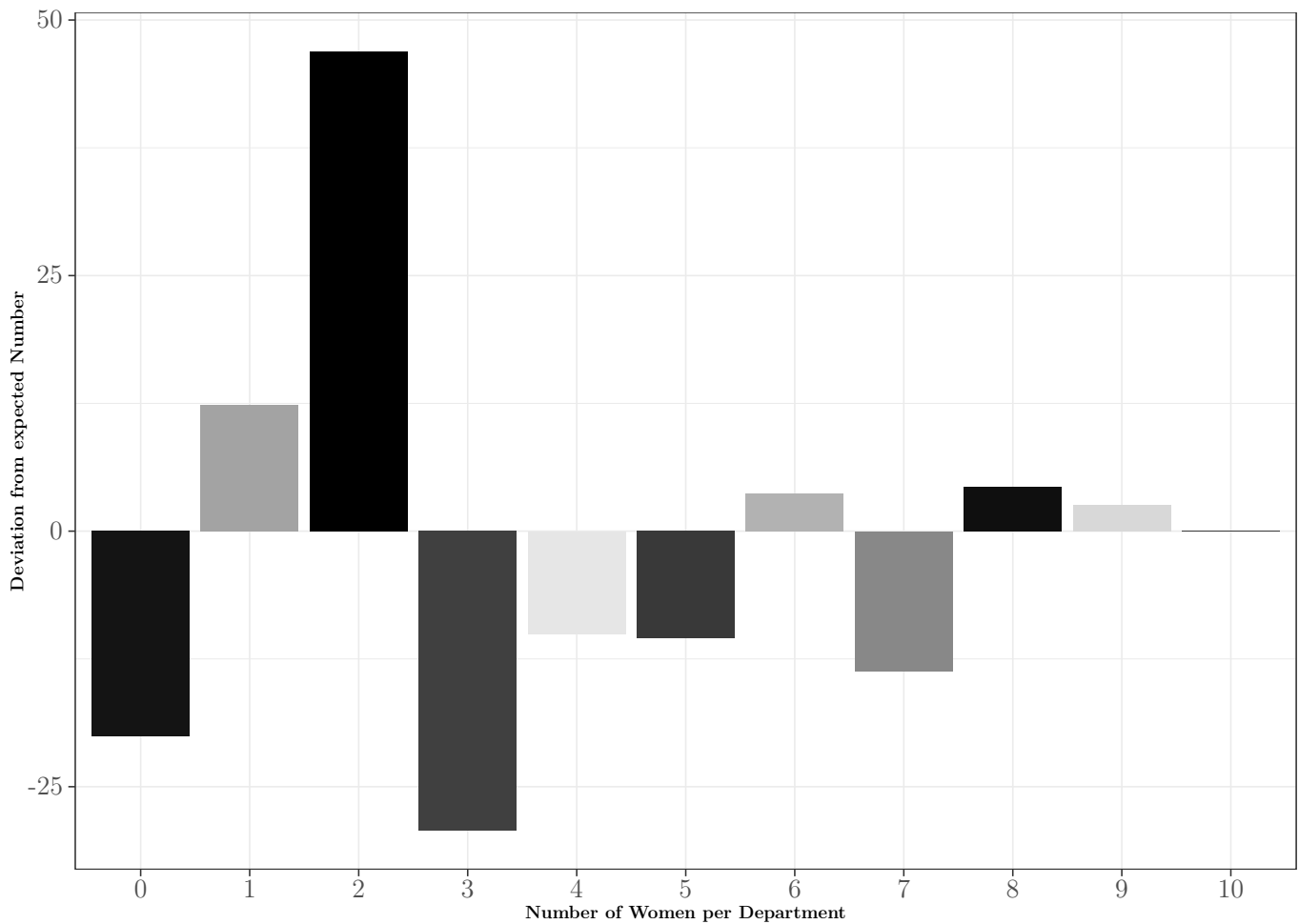
\begin{figure}
\hspace{-1.5cm}
\begin{tikzpicture}[x=1pt,y=1pt]
\definecolor{fillColor}{RGB}{255,255,255}
\path[use as bounding box,fill=fillColor] (0,0) rectangle (520.34,397.48);
\begin{scope}
\path[clip] (  0.00,  0.00) rectangle (520.34,397.48);
\definecolor{drawColor}{RGB}{255,255,255}

\path[draw=drawColor,line width= 0.6pt,line join=round,line cap=round,fill=fillColor] (  0.00,  0.00) rectangle (520.34,397.48);
\end{scope}
\begin{scope}
\path[clip] ( 30.24, 26.28) rectangle (514.84,366.83);
\definecolor{fillColor}{RGB}{255,255,255}

\path[fill=fillColor] ( 30.24, 26.28) rectangle (514.84,366.83);
\definecolor{drawColor}{gray}{0.92}

\path[draw=drawColor,line width= 0.3pt,line join=round] ( 30.24,109.90) --
	(514.84,109.90);

\path[draw=drawColor,line width= 0.3pt,line join=round] ( 30.24,211.45) --
	(514.84,211.45);

\path[draw=drawColor,line width= 0.3pt,line join=round] ( 30.24,313.01) --
	(514.84,313.01);

\path[draw=drawColor,line width= 0.6pt,line join=round] ( 30.24, 59.12) --
	(514.84, 59.12);

\path[draw=drawColor,line width= 0.6pt,line join=round] ( 30.24,160.68) --
	(514.84,160.68);

\path[draw=drawColor,line width= 0.6pt,line join=round] ( 30.24,262.23) --
	(514.84,262.23);

\path[draw=drawColor,line width= 0.6pt,line join=round] ( 30.24,363.79) --
	(514.84,363.79);

\path[draw=drawColor,line width= 0.6pt,line join=round] ( 56.20, 26.28) --
	( 56.20,366.83);

\path[draw=drawColor,line width= 0.6pt,line join=round] ( 99.47, 26.28) --
	( 99.47,366.83);

\path[draw=drawColor,line width= 0.6pt,line join=round] (142.74, 26.28) --
	(142.74,366.83);

\path[draw=drawColor,line width= 0.6pt,line join=round] (186.00, 26.28) --
	(186.00,366.83);

\path[draw=drawColor,line width= 0.6pt,line join=round] (229.27, 26.28) --
	(229.27,366.83);

\path[draw=drawColor,line width= 0.6pt,line join=round] (272.54, 26.28) --
	(272.54,366.83);

\path[draw=drawColor,line width= 0.6pt,line join=round] (315.81, 26.28) --
	(315.81,366.83);

\path[draw=drawColor,line width= 0.6pt,line join=round] (359.08, 26.28) --
	(359.08,366.83);

\path[draw=drawColor,line width= 0.6pt,line join=round] (402.35, 26.28) --
	(402.35,366.83);

\path[draw=drawColor,line width= 0.6pt,line join=round] (445.61, 26.28) --
	(445.61,366.83);

\path[draw=drawColor,line width= 0.6pt,line join=round] (488.88, 26.28) --
	(488.88,366.83);
\definecolor{fillColor}{gray}{0.08}

\path[fill=fillColor] ( 36.73, 79.09) rectangle ( 75.67,160.68);
\definecolor{fillColor}{gray}{0.64}

\path[fill=fillColor] ( 80.00,160.68) rectangle (118.94,210.83);
\definecolor{fillColor}{RGB}{0,0,0}

\path[fill=fillColor] (123.26,160.68) rectangle (162.21,351.35);
\definecolor{fillColor}{gray}{0.25}

\path[fill=fillColor] (166.53, 41.76) rectangle (205.47,160.68);
\definecolor{fillColor}{gray}{0.90}

\path[fill=fillColor] (209.80,119.80) rectangle (248.74,160.68);
\definecolor{fillColor}{RGB}{57,57,57}

\path[fill=fillColor] (253.07,118.18) rectangle (292.01,160.68);
\definecolor{fillColor}{gray}{0.70}

\path[fill=fillColor] (296.34,160.68) rectangle (335.28,175.57);
\definecolor{fillColor}{RGB}{136,136,136}

\path[fill=fillColor] (339.61,105.04) rectangle (378.55,160.68);
\definecolor{fillColor}{gray}{0.06}

\path[fill=fillColor] (382.88,160.68) rectangle (421.82,178.29);
\definecolor{fillColor}{RGB}{216,216,216}

\path[fill=fillColor] (426.14,160.68) rectangle (465.09,171.19);
\definecolor{fillColor}{RGB}{58,58,58}

\path[fill=fillColor] (469.41,160.43) rectangle (508.35,160.68);
\definecolor{drawColor}{gray}{0.20}

\path[draw=drawColor,line width= 0.6pt,line join=round,line cap=round] ( 30.24, 26.28) rectangle (514.84,366.83);
\end{scope}
\begin{scope}
\path[clip] (  0.00,  0.00) rectangle (520.34,397.48);
\definecolor{drawColor}{gray}{0.30}

\node[text=drawColor,anchor=base east,inner sep=0pt, outer sep=0pt, scale=  0.88] at ( 25.29, 56.09) {-25};

\node[text=drawColor,anchor=base east,inner sep=0pt, outer sep=0pt, scale=  0.88] at ( 25.29,157.65) {0};

\node[text=drawColor,anchor=base east,inner sep=0pt, outer sep=0pt, scale=  0.88] at ( 25.29,259.20) {25};

\node[text=drawColor,anchor=base east,inner sep=0pt, outer sep=0pt, scale=  0.88] at ( 25.29,360.76) {50};
\end{scope}
\begin{scope}
\path[clip] (  0.00,  0.00) rectangle (520.34,397.48);
\definecolor{drawColor}{gray}{0.20}

\path[draw=drawColor,line width= 0.6pt,line join=round] ( 27.49, 59.12) --
	( 30.24, 59.12);

\path[draw=drawColor,line width= 0.6pt,line join=round] ( 27.49,160.68) --
	( 30.24,160.68);

\path[draw=drawColor,line width= 0.6pt,line join=round] ( 27.49,262.23) --
	( 30.24,262.23);

\path[draw=drawColor,line width= 0.6pt,line join=round] ( 27.49,363.79) --
	( 30.24,363.79);
\end{scope}
\begin{scope}
\path[clip] (  0.00,  0.00) rectangle (520.34,397.48);
\definecolor{drawColor}{gray}{0.20}

\path[draw=drawColor,line width= 0.6pt,line join=round] ( 56.20, 23.53) --
	( 56.20, 26.28);

\path[draw=drawColor,line width= 0.6pt,line join=round] ( 99.47, 23.53) --
	( 99.47, 26.28);

\path[draw=drawColor,line width= 0.6pt,line join=round] (142.74, 23.53) --
	(142.74, 26.28);

\path[draw=drawColor,line width= 0.6pt,line join=round] (186.00, 23.53) --
	(186.00, 26.28);

\path[draw=drawColor,line width= 0.6pt,line join=round] (229.27, 23.53) --
	(229.27, 26.28);

\path[draw=drawColor,line width= 0.6pt,line join=round] (272.54, 23.53) --
	(272.54, 26.28);

\path[draw=drawColor,line width= 0.6pt,line join=round] (315.81, 23.53) --
	(315.81, 26.28);

\path[draw=drawColor,line width= 0.6pt,line join=round] (359.08, 23.53) --
	(359.08, 26.28);

\path[draw=drawColor,line width= 0.6pt,line join=round] (402.35, 23.53) --
	(402.35, 26.28);

\path[draw=drawColor,line width= 0.6pt,line join=round] (445.61, 23.53) --
	(445.61, 26.28);

\path[draw=drawColor,line width= 0.6pt,line join=round] (488.88, 23.53) --
	(488.88, 26.28);
\end{scope}
\begin{scope}
\path[clip] (  0.00,  0.00) rectangle (520.34,397.48);
\definecolor{drawColor}{gray}{0.30}

\node[text=drawColor,anchor=base,inner sep=0pt, outer sep=0pt, scale=  0.88] at ( 56.20, 15.27) {0};

\node[text=drawColor,anchor=base,inner sep=0pt, outer sep=0pt, scale=  0.88] at ( 99.47, 15.27) {1};

\node[text=drawColor,anchor=base,inner sep=0pt, outer sep=0pt, scale=  0.88] at (142.74, 15.27) {2};

\node[text=drawColor,anchor=base,inner sep=0pt, outer sep=0pt, scale=  0.88] at (186.00, 15.27) {3};

\node[text=drawColor,anchor=base,inner sep=0pt, outer sep=0pt, scale=  0.88] at (229.27, 15.27) {4};

\node[text=drawColor,anchor=base,inner sep=0pt, outer sep=0pt, scale=  0.88] at (272.54, 15.27) {5};

\node[text=drawColor,anchor=base,inner sep=0pt, outer sep=0pt, scale=  0.88] at (315.81, 15.27) {6};

\node[text=drawColor,anchor=base,inner sep=0pt, outer sep=0pt, scale=  0.88] at (359.08, 15.27) {7};

\node[text=drawColor,anchor=base,inner sep=0pt, outer sep=0pt, scale=  0.88] at (402.35, 15.27) {8};

\node[text=drawColor,anchor=base,inner sep=0pt, outer sep=0pt, scale=  0.88] at (445.61, 15.27) {9};

\node[text=drawColor,anchor=base,inner sep=0pt, outer sep=0pt, scale=  0.88] at (488.88, 15.27) {10};
\end{scope}
\begin{scope}
\path[clip] (  0.00,  0.00) rectangle (520.34,397.48);
\definecolor{drawColor}{RGB}{0,0,0}

\node[text=drawColor,anchor=base,inner sep=0pt, outer sep=0pt, scale=  0.60] at (272.54,  6.67) {\bfseries Number of Women per Department};
\end{scope}
\begin{scope}
\path[clip] (  0.00,  0.00) rectangle (520.34,397.48);
\definecolor{drawColor}{RGB}{0,0,0}

\node[text=drawColor,rotate= 90.00,anchor=base,inner sep=0pt, outer sep=0pt, scale=  0.60] at (  9.64,196.56) {\bfseries Deviation from expected Number};
\end{scope}
\begin{scope}
\path[clip] (  0.00,  0.00) rectangle (520.34,397.48);
\definecolor{drawColor}{RGB}{0,0,0}

\node[text=drawColor,anchor=base west,inner sep=0pt, outer sep=0pt, scale=  0.70] at ( 30.24,373.69) {\bfseries Mean over all 50 disciplines, shaded by p-value, $p_{min}=0.011,$ $p_{max}=0.27$};
\end{scope}
\begin{scope}
\path[clip] (  0.00,  0.00) rectangle (520.34,397.48);
\definecolor{drawColor}{RGB}{0,0,0}

\node[text=drawColor,anchor=base west,inner sep=0pt, outer sep=0pt, scale=  0.90] at ( 30.24,385.77) {\bfseries Deviation from expected Number of Departments with zero to 10 women};
\end{scope}
\end{tikzpicture}
		\caption{Deviation from expected value (in absolute numbers), shaded by joint P-Value, the darker the shading, the lower the p-value. Lowest p-value corresponds to the deviation from exactly 2 women departments with a p-value of $0.011$.  }\label{fig:res_uc}
	\end{figure}

\begin{figure}[H]
\hspace{-1.5cm}
		\input{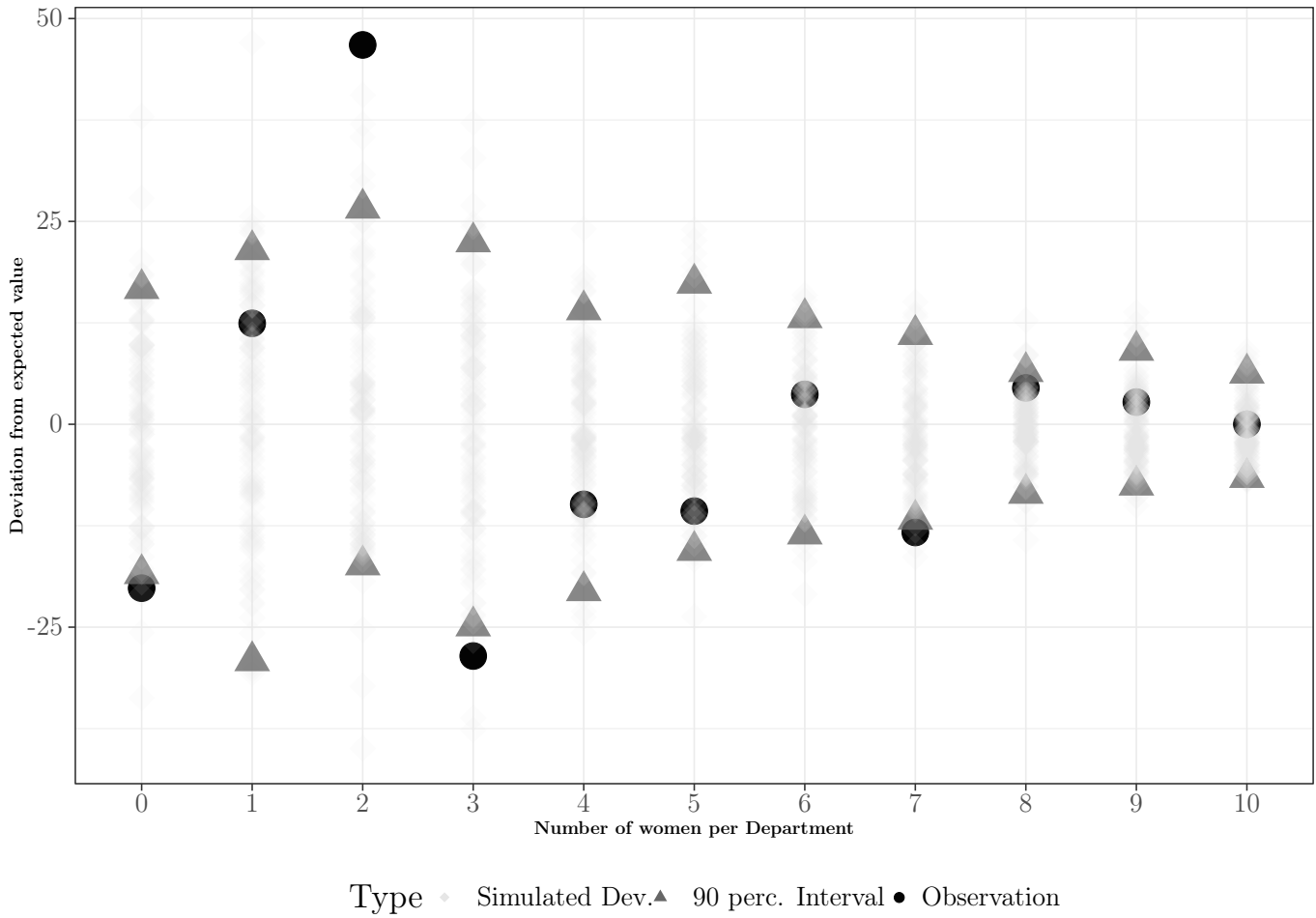}
	
		\caption{Deviation from expected value summed over all disciplines and divided by the absolute number of disciplines. }\label{fig:res_uc_boot}
	\end{figure}

The top panel of Figure (\ref{fig:res_uc}) graphically shows the summed deviations ($B(z)=\sum_{1=d}^{n=1737}B_d(z)$) of the number of departments with exactly $z$ women, for $z=0,1,...,10$, summed over all departments, shaded by the p-values with the critical values from a two-sided test from a standard normal distribution. Darker shades indicate lower p-values. The deviations and p-values can also be found in Table (\ref{table_p_val}). 
The results indicate that the joint hypothesis of gender-blind hiring can be rejected for several values of $z$. Specifically, there are too few departments with exactly zero women (there are about 20 departments less than expected with an associated p-value of 0.03) and too few departments with three women (with p-value 0.09), four women (p-value 0.28, five women (p-value 0.08) and seven women (p-value 0.18). On the other hand, there are about 12 departments too many with precisely one woman (p-value 0.21) and 47 too many departments with two women (p-value 0.01). This indicates a two-women implicit quota and that departments specifically try to avoid having no female professors. 
As an alternative to the test above, I depict the results from the parametric bootstrap procedure described in Section (\ref{sec:boot}) in Figure (\ref{fig:res_uc_boot}). The transparent diamonds represent individual bootstrap draws $B(z)^{b^*}$, the grey triangles represent the empirical 90\% interval, and the black dots are the observed ${B}(z)$ from the data. As already stated, the results are generally in line with the results from the test in terms of the empirical p-values. Both the observed deviations for zero women, two and three women lie unambiguously outside the 90\% interval. Therefore, the results from the bootstrap procedure confirm the presence of a two-women department implicit quota and an aversion against zero women in the department.

When I compare the two methods, the bootstrap implies that for larger $z$, there is more uncertainty than the asymptotic approximation would suggest. Intuitively this makes sense, as with the overall small $p_s$, we would only expect very few departments with \emph{many} women. Therefore even minor deviations (because actual observations are integers and expected values are not) would be significant compared to the normal distribution. There exist continuity adjustments for normal approximations of the binomial distribution (see, e.g., \cite{feller2015normal}), and this could probably be applied in this situation if one wanted to avoid the parametric bootstrap procedure. 

 \subsection{Independence and constant hiring probabilities}\label{sec:ind}
 
 Both the asymptotic normality result and the parametric bootstrap rely on the independence of hiring decisions across departments, both within and across disciplines. This assumption is usually justified in this context, as the hiring of professors is essentially sampling with replacement, i.e., the success probability is approximately constant, and knowledge of the outcome of a previous hiring decision at a different department should not influence the gender of the next hire. However, I also assume that hiring probabilities are constant across departments and as it is not possible to hire the same \emph{person} twice, this assumption might be problematic. With the overall number of departments and professors large, the change in the overall probability of hiring a woman is $\approx$ unchanged when the women already in ``own-department'' do not count towards the hiring probability. To verify that this is a reasonable assumption in this application, I calculate the ``leave-one-out'' probability $p_{d|s}^{loo}$ of hiring a woman and how $p_{d|s}^{loo}$ differ from the overall probability $p_s$. Table (\ref{table_sd_llo}) in the Appendix shows the standard deviation of the leave-one-out probabilities within individual disciplines. They are all (very) small and only for 0.041 percent of departments (with 1737 departments from 50 different disciplines) is the t-test for equality of means of the ``leave-one-out'' mean and the overall mean rejected with $\alpha=0.05$. Thus, the constant probability assumption appears to be justified in this context.

\newpage

\begin{table}[h]
\begin{center}
\sisetup{parse-numbers=false, table-text-alignment=right}
\begin{tabular}{l r S[table-format=1.3]}
\toprule
 & {$B(z)$} & {p-value} \\
\midrule
0  & -20.080 & 0.031 \\
1  & 12.350  & 0.212 \\
2  & 46.940  & 0.012 \\
3  & -29.270 & 0.092 \\
4  & -10.060 & 0.283 \\
5  & -10.460 & 0.083 \\
6  & 3.670   & 0.230 \\
7  & -13.690 & 0.181 \\
8  & 4.340   & 0.025 \\
9  & 2.590   & 0.270 \\
10 & -0.060  & 0.084 \\
\bottomrule
\multicolumn{3}{l}{}
\end{tabular}
\caption{Summed Deviations and p-values the asymptotic test in \eqref{eq:test_Hz_II}. }
\label{table_p_val}
\end{center}
\end{table}


%

\newpage

\subsection{Effects of a fictional quota}\label{sec:fic_quota} 
\begin{figure}[H]
		\input{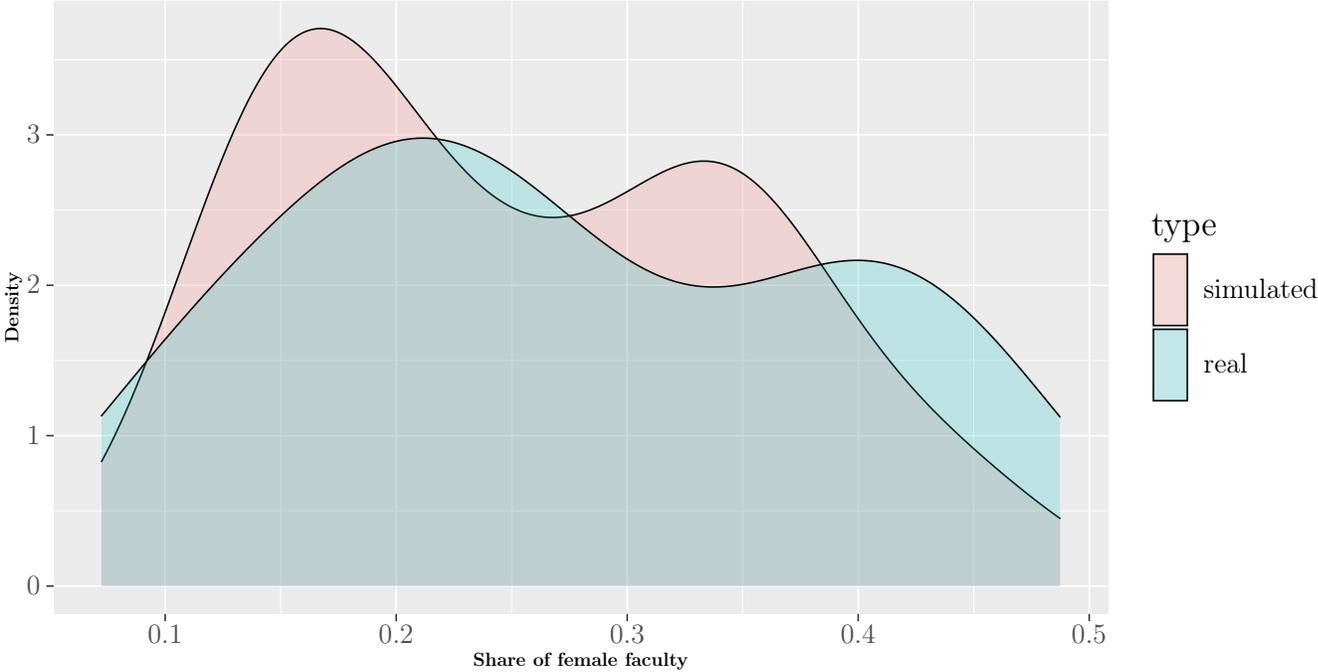}
		\caption{Density plot for the distribution of real and simulated shares of female faculty across disciplines.}\label{fig:share_sim}
	\end{figure}

Given the analysis above, we might speculate about implicit quotas in specific disciplines or we might hypothesize that disciplines that have a low share of female professors provide a more hostile environment and therefore are more likely to discriminate against women. However, the correlation $\rho_z$ between having a negative deviation from the expected number of zero women departments and the share of women is $\rho_0=-0.035$ (p-value $0.86$) and $\rho_3=0.07$ (p-value $0.62$). Equally, there is no clear correlation with being a STEM/non-STEM discipline. These results suggest that the type of discipline with an implicit quota is not easily categorized. This result might seem puzzling at first glance, but when considering the bi-modal distribution of average department sizes, it becomes clear that implementing a quota would have heterogeneous effects on the share of women within the discipline, depending on the average department size. Most of the non-STEM fields (with the notable exceptions economics and law) which have the largest share of female faculty have average department sizes below ten professors: this implies that a two women quota would mechanically lead to a much higher female share in these disciplines than in mathematics or economics, where the average department size is much larger. This exercise shows that we cannot rule out that non-STEM disciplines in the humanities have a larger share of female professors because an implicit quota has a heterogeneous effect depending on average department sizes in the discipline rather than discipline-specific characteristics. To illustrate this point, consider Figure (\ref{fig:share_sim}): here I simulate the hypothetical shares of women within subjects that would result from taking the vector of real department sizes $k_s$ for all disciplines $s$ and imposing a two women quota per department, as shown for an example in Table (\ref{tab_share}) for Economics. 
\begin{table}[H]
\centering
\begin{tabular}{rrr}
  \hline
Subject & Size & Fe. Faculty \\ 
  \hline
Econ & 6 & 0 \\ 
  Econ & 16 & 2 \\ 
  Econ & 12 & 2 \\ 
  Econ & 7 & 1 \\ 
  Econ & 23 & 5 \\ 
  $\vdots$&$\vdots$&$\vdots$\\
   \hline
\end{tabular}
\hspace{1cm}
\begin{tabular}{rrr}
  \hline
Code & Size & FE Faculty Quota  \\ 
  \hline
Econ & 6 & 2 \\ 
  Econ & 16 & 2 \\ 
  Econ & 12 & 2 \\ 
  Econ & 7 & 2 \\ 
  Econ & 23 & 2 \\ 
   $\vdots$&$\vdots$&$\vdots$\\
   \hline
\end{tabular}\caption{Left are the department size and the real number of female faculty, the right panel shows the actual number replaced by the shares.}\label{tab_share}
\end{table}

\newpage

\newpage

A two-women quota per department would lead to a mean female share of $\mu^s_f= 0.254$, instead of the actual share $\mu_f= 0.257$. It can reproduce the bimodal distribution of the female share distribution, although the peaks are slightly shifted to the left. This exercise does not intend to state that this type of quota is consciously implemented, but that a certain adherence to ``discipline-standard'' hiring may lead to a distribution of female shares very similar to the one we observe. 

\newpage

\newpage



%



\doublespacing
\section{Conclusion and Discussion}\label{sec:disc}

This paper introduced a tool to detect an under-researched form of hiring bias: implicit quotas. I derive a test under the null of gender-blind hiring that requires no additional information about individual hires under some assumptions. I derive the asymptotic distribution of this test statistic and, as an alternative, propose a numerical approximation of the exact distribution which is known, but infeasible to calculate in most cases. This test can be used to analyze a variety of other hiring settings. 

  Analyzing employment patterns of female professors across departments and disciplines allows me to show that there exists a one- to two-women quota on the department level. This type of discrimination or hiring bias cannot be explained with traditional explanations for discrimination, for example either pure statistical or taste-based discrimination.\footnote{\cite{bayer2016diversity} categorize research findings in economics into supply-side factors (i.e. hostile work environments,  (\cite{wu2018gendered}) and demands side factors (such as  a lower probability to get published or accepted at conferences (conditional on paper quality and a range of controls) (\cite{card2020referees}, \cite{hospido2019gender}, higher discounting of women's contributions when working with male co-authors (\cite{sarsons2017recognition}, more ``unfair'' questions during seminars (\cite{dupas2020gender} or lower teaching evaluations (\cite{boring2017gender}, \cite{mengel2019gender}). Neither of these is a sufficient explanation for the presence of implicit quotas. } 
 I also show that this type of implicit quota can potentially explain part of the gap in the female professor share among disciplines. The fact that I can distinguish between the implicit quota being centered around the mean (i.e., around the relative share of female professors in a discipline, which is not the case) and a specific discrete number poses exciting questions that I have not seen addressed before, namely how humans perceive structures that have the ``correct level'' of diversity. 

\bigskip

It is important to emphasize that the observed patterns described above only allow me to make statements about whether or not the observed distribution is gender-blind: we cannot say whether the ``real/natural'' proportion of female professors $p_s$ (without the implicit quota) would be larger, smaller or precisely the same. Furthermore, direct discrimination in implicit quotas (which leads to discrimination in both directions), gender differences in personal attributes, preferences, and implicit biases are not mutually exclusive. There is even a plausible link between biases and direct discrimination; as \cite{goldin2014pollution} argues, if men perceive women as less qualified and care about the prestige of their profession, they may fear that a woman's entering a profession signals a change in the prestige of the profession and therefore might block entry to otherwise qualified women.

\bigskip

The analysis in this paper is sufficient to show that hiring is not gender blind. My work, therefore, has several implications for both further research and policy. 

 Concerning research, any ex-post analysis of performance measures and researcher productivity should be conducted only with the caveat in mind that there might be implicit quotas. Implicit quotas could potentially lead to substantially more quality \emph{heterogeneity} among the underrepresented groups that might be obscured when researchers only consider averages. 
 
 Concerning policy, my results imply that funders and organizations that engage in diversity pressure without explicit targets should be conscious of the potential for implicit quotas and carefully monitor progress. My results also imply that if policymakers have a specific diversity target in mind, then an explicit quota might be more effective, as implicit quotas might incur the exact costs as explicit quotas while not achieving the same level of progress.


\clearpage

 \clearpage
\bibliographystyle{ecta}
\bibliography{biblio}

\begin{thebibliography}{29}
\newcommand{\enquote}[1]{``#1''}
\expandafter\ifx\csname natexlab\endcsname\relax\def\natexlab#1{#1}\fi

\bibitem[\protect\citeauthoryear{AllBright~Stiftung}{AllBright~Stiftung}{2020}]{allbright2020}
\textsc{AllBright~Stiftung, B.} (2020): \enquote{AllBright Bericht: Deutscher
  Sonderweg: Frauenanteil in DAX-Vorständen sinkt in der Krise,} .

\bibitem[\protect\citeauthoryear{Arcidiacono, Kinsler, and Ransom}{Arcidiacono
  et~al.}{2020}]{Arciharvardnber}
\textsc{Arcidiacono, P., J.~Kinsler, and T.~Ransom} (2020): \enquote{Asian
  American Discrimination in Harvard Admissions,} \emph{NBER Working Paper}.

\bibitem[\protect\citeauthoryear{Auriol, Friebel, and Wilhelm}{Auriol
  et~al.}{2020}]{friebel2019women}
\textsc{Auriol, E., G.~Friebel, and S.~Wilhelm} (2020): \enquote{Women in
  European Economics,} \emph{Women in Economics, edited by S. Lundberg},
  26--30.

\bibitem[\protect\citeauthoryear{Bagues and Campa}{Bagues and
  Campa}{2021}]{bagues2021can}
\textsc{Bagues, M. and P.~Campa} (2021): \enquote{Can gender quotas in
  candidate lists empower women? Evidence from a regression discontinuity
  design,} \emph{Journal of Public Economics}, 194, 104315.

\bibitem[\protect\citeauthoryear{Balafoutas and Sutter}{Balafoutas and
  Sutter}{2012}]{balafoutas2012affirmative}
\textsc{Balafoutas, L. and M.~Sutter} (2012): \enquote{Affirmative action
  policies promote women and do not harm efficiency in the laboratory,}
  \emph{Science}, 335, 579--582.

\bibitem[\protect\citeauthoryear{Bayer and Rouse}{Bayer and
  Rouse}{2016}]{bayer2016diversity}
\textsc{Bayer, A. and C.~E. Rouse} (2016): \enquote{Diversity in the economics
  profession: A new attack on an old problem,} \emph{Journal of Economic
  Perspectives}, 30, 221--42.

\bibitem[\protect\citeauthoryear{Beaman, Duflo, Pande, and Topalova}{Beaman
  et~al.}{2012}]{beaman2012female}
\textsc{Beaman, L., E.~Duflo, R.~Pande, and P.~Topalova} (2012):
  \enquote{Female leadership raises aspirations and educational attainment for
  girls: A policy experiment in India,} \emph{science}, 335, 582--586.

\bibitem[\protect\citeauthoryear{Bertrand, Black, Jensen, and
  Lleras-Muney}{Bertrand et~al.}{2018}]{bertrand2018breaking}
\textsc{Bertrand, M., S.~E. Black, S.~Jensen, and A.~Lleras-Muney} (2018):
  \enquote{Breaking the glass ceiling? The effect of board quotas on female
  labour market outcomes in Norway,} \emph{The Review of Economic Studies}, 86,
  191--239.

\bibitem[\protect\citeauthoryear{Boring}{Boring}{2017}]{boring2017gender}
\textsc{Boring, A.} (2017): \enquote{Gender biases in student evaluations of
  teaching,} \emph{Journal of public economics}, 145, 27--41.

\bibitem[\protect\citeauthoryear{Buser, Niederle, and Oosterbeek}{Buser
  et~al.}{2014}]{buser2014gender}
\textsc{Buser, T., M.~Niederle, and H.~Oosterbeek} (2014): \enquote{Gender,
  competitiveness, and career choices,} \emph{The Quarterly Journal of
  Economics}, 129, 1409--1447.

\bibitem[\protect\citeauthoryear{Card, DellaVigna, Funk, and Iriberri}{Card
  et~al.}{2020}]{card2020referees}
\textsc{Card, D., S.~DellaVigna, P.~Funk, and N.~Iriberri} (2020): \enquote{Are
  Referees and Editors in Economics Gender Neutral?} \emph{The Quarterly
  Journal of Economics}, 135, 269--327.

\bibitem[\protect\citeauthoryear{Chang, Milkman, Chugh, and Akinola}{Chang
  et~al.}{2019}]{chang2019diversity}
\textsc{Chang, E.~H., K.~L. Milkman, D.~Chugh, and M.~Akinola} (2019):
  \enquote{Diversity thresholds: How social norms, visibility, and scrutiny
  relate to group composition,} \emph{Academy of Management Journal}, 62,
  144--171.

\bibitem[\protect\citeauthoryear{Dupas, Modestino, Niederle, and Wolfers}{Dupas
  et~al.}{2020}]{dupas2020gender}
\textsc{Dupas, P., A.~Modestino, M.~Niederle, and J.~Wolfers} (2020):
  \enquote{Gender and the Dynamics of Economics Seminars,} Tech. rep.

\bibitem[\protect\citeauthoryear{Eurostat}{Eurostat}{2020{\natexlab{a}}}]{eurostat2019}
\textsc{Eurostat} (2020{\natexlab{a}}): \enquote{Gender Statistics Database,} .

\bibitem[\protect\citeauthoryear{Eurostat}{Eurostat}{2020{\natexlab{b}}}]{eurostat20192}
---\hspace{-.1pt}---\hspace{-.1pt}--- (2020{\natexlab{b}}): \enquote{Gender
  Statistics Database,} .

\bibitem[\protect\citeauthoryear{for Human~Rights}{for
  Human~Rights}{2020}]{eindhoven2020}
\textsc{for Human~Rights, T. N.~I.} (2020): \enquote{Advies aan minister Dekker
  voor Rechtsbescherming over het evenwichtiger maken van de man/vrouw
  verhouding vennootschappen,} .

\bibitem[\protect\citeauthoryear{Goldin}{Goldin}{2014}]{goldin2014pollution}
\textsc{Goldin, C.} (2014): \enquote{A pollution theory of discrimination: male
  and female differences in occupations and earnings,} in \emph{Human capital
  in history: The American record}, University of Chicago Press, 313--348.

\bibitem[\protect\citeauthoryear{Hong}{Hong}{2013}]{hong2013computing}
\textsc{Hong, Y.} (2013): \enquote{On computing the distribution function for
  the Poisson binomial distribution,} \emph{Computational Statistics \& Data
  Analysis}, 59, 41--51.

\bibitem[\protect\citeauthoryear{Hospido and Sanz}{Hospido and
  Sanz}{2019}]{hospido2019gender}
\textsc{Hospido, L. and C.~Sanz} (2019): \enquote{Gender gaps in the evaluation
  of research: evidence from submissions to economics conferences,} \emph{IZA
  Discussion Paper}.

\bibitem[\protect\citeauthoryear{Huang, Gates, Sinatra, and Barab{\'a}si}{Huang
  et~al.}{2020}]{huang2020historical}
\textsc{Huang, J., A.~J. Gates, R.~Sinatra, and A.-L. Barab{\'a}si} (2020):
  \enquote{Historical comparison of gender inequality in scientific careers
  across countries and disciplines,} \emph{Proceedings of the National Academy
  of Sciences}, 117, 4609--4616.

\bibitem[\protect\citeauthoryear{Lehmann}{Lehmann}{2004}]{lehmann2004elements}
\textsc{Lehmann, E.~L.} (2004): \emph{Elements of large-sample theory},
  Springer Science \& Business Media.

\bibitem[\protect\citeauthoryear{Liu and Quertermous}{Liu and
  Quertermous}{2018}]{liu2018approximating}
\textsc{Liu, B. and T.~Quertermous} (2018): \enquote{Approximating the Sum of
  Independent Non-Identical Binomial Random Variables.} \emph{R Journal}, 10.

\bibitem[\protect\citeauthoryear{Lundberg and Stearns}{Lundberg and
  Stearns}{2019}]{lundberg2020}
\textsc{Lundberg, S. and J.~Stearns} (2019): \enquote{Women in economics:
  Stalled progress,} \emph{Journal of Economic Perspectives}, 33, 3--22.

\bibitem[\protect\citeauthoryear{Maida and Weber}{Maida and
  Weber}{2019}]{maida2019female}
\textsc{Maida, A. and A.~Weber} (2019): \enquote{Female Leadership and Gender
  Gap within Firms: Evidence from an Italian Board Reform,} \emph{IZA
  Discussion Paper}.

\bibitem[\protect\citeauthoryear{Mengel, Sauermann, and Z{\"o}litz}{Mengel
  et~al.}{2019}]{mengel2019gender}
\textsc{Mengel, F., J.~Sauermann, and U.~Z{\"o}litz} (2019): \enquote{Gender
  bias in teaching evaluations,} \emph{Journal of the European Economic
  Association}, 17, 535--566.

\bibitem[\protect\citeauthoryear{Paryavi, Bohnet, and van Geen}{Paryavi
  et~al.}{2019}]{paryavi2019descriptive}
\textsc{Paryavi, M., I.~Bohnet, and A.~van Geen} (2019): \enquote{Descriptive
  norms and gender diversity: Reactance from men,} 2.

\bibitem[\protect\citeauthoryear{Porter and Serra}{Porter and
  Serra}{2020}]{porter2020gender}
\textsc{Porter, C. and D.~Serra} (2020): \enquote{Gender differences in the
  choice of major: The importance of female role models,} \emph{American
  Economic Journal: Applied Economics}, 12, 226--54.

\bibitem[\protect\citeauthoryear{Sarsons}{Sarsons}{2017}]{sarsons2017recognition}
\textsc{Sarsons, H.} (2017): \enquote{Recognition for group work: Gender
  differences in academia,} \emph{American Economic Review}, 107, 141--45.

\bibitem[\protect\citeauthoryear{Wu}{Wu}{2018}]{wu2018gendered}
\textsc{Wu, A.~H.} (2018): \enquote{Gendered language on the economics job
  market rumors forum,} \emph{AEA Papers and Proceedings}, 108, 175--79.

\end{thebibliography}

\section*{Appendix}

\begin{table}[h]
\scriptsize
\begin{center}
\begin{tabular}{l >\boldmath c c}
\toprule
Discipline & Standard Deviation, Leave-one-out Shares & Shares \\
\midrule
Humanities                                  & $0.0094$ & $0.4006$ \\
Theology (protestant)                       & $0.0059$ & $0.2308$ \\
Theology (catholic)                         & $0.0049$ & $0.1924$ \\
Philosophy                                  & $0.0029$ & $0.1616$ \\
History                                     & $0.0029$ & $0.3068$ \\
Literature and comparative language studies & $0.0114$ & $0.4360$ \\
German Studies                              & $0.0026$ & $0.4185$ \\
English and American Studies                & $0.0034$ & $0.4393$ \\
Roman Language Studies                      & $0.0062$ & $0.3985$ \\
Other Language Studies             & $0.0073$ & $0.4100$ \\
Cultural Studies                            & $0.0115$ & $0.4874$ \\
Sports                                      & $0.0050$ & $0.2039$ \\
Law, Economics and Social Sciences & $0.0095$ & $0.2844$ \\
Political Sciences                          & $0.0042$ & $0.2565$ \\
Sociology                                   & $0.0027$ & $0.3636$ \\
Social Work                                 & $0.0180$ & $0.4702$ \\
Law                                         & $0.0022$ & $0.1614$ \\
Economics                                   & $0.0012$ & $0.1838$ \\
Psychology                                  & $0.0027$ & $0.3919$ \\
Paedagogy                                   & $0.0024$ & $0.4400$ \\
STEM, general                               & $0.0402$ & $0.2540$ \\
Mathematics                                 & $0.0016$ & $0.1709$ \\
Physics and Astronomy                       & $0.0012$ & $0.0976$ \\
Chemistry                                   & $0.0015$ & $0.1446$ \\
Pharmacy                                    & $0.0060$ & $0.2187$ \\
Biology                                     & $0.0020$ & $0.2252$ \\
Geology                                     & $0.0025$ & $0.1284$ \\
Geography                                   & $0.0035$ & $0.2434$ \\
Health Sciences, general                    & $0.0196$ & $0.4773$ \\
Preclinical Medicine                        & $0.0035$ & $0.1715$ \\
Clinical-theoretical Medicine               & $0.0027$ & $0.1775$ \\
Clinical-Practical Medicine                 & $0.0021$ & $0.1236$ \\
Clinical-theoretical Veterinary Science     & $0.0221$ & $0.2380$ \\
Clinical-practical Veterinary Science       & $0.0361$ & $0.2962$ \\
Agricultural Science                        & $0.0078$ & $0.2013$ \\
Dietary Studies                             & $0.0247$ & $0.3638$ \\
Mechanical Engineering                      & $0.0016$ & $0.0887$ \\
Electrical Engineering                      & $0.0024$ & $0.0724$ \\
Architecture                                & $0.0058$ & $0.2476$ \\
Spatial Planning                            & $0.0031$ & $0.3022$ \\
Civil Engineering                           & $0.0040$ & $0.0961$ \\
Computer Science                            & $0.0012$ & $0.1062$ \\
Material Science                            & $0.0102$ & $0.1991$ \\
Art, general                                & $0.0046$ & $0.4422$ \\
Visual Arts                                 & $0.0043$ & $0.3547$ \\
Design                                      & $0.0116$ & $0.3624$ \\
Performing Arts                             & $0.0084$ & $0.3904$ \\
Music                                       & $0.0024$ & $0.2315$ \\
Central scientific Departments              & $0.0072$ & $0.3092$ \\
Institutes associated with Universities     & $0.0430$ & $0.2781$ \\
\bottomrule
\multicolumn{3}{l}{\scriptsize{$^{***}p<0.001$; $^{**}p<0.01$; $^{*}p<0.05$}}
\end{tabular}
\caption{Standard deviation leave-one-out means by discipline}
\label{table_sd_llo}
\end{center}
\end{table}

\end{document}